\documentclass[traditabstract]{aa} 

\usepackage[pdftex,final]{graphicx}  
\usepackage[comma,authoryear]{natbib} 
\bibpunct{(}{)}{,}{a}{}{,}            
\usepackage{amssymb}

\begin{document}

   \title{Gravity modes in rapidly rotating polytropic stars}


   \author{J. Ballot\inst{1}
               \and
           F. Ligni\`eres\inst{1}
               \and
               D. R. Reese\inst{2}
               \and
               M. Rieutord\inst{1}
          }

   \institute{Laboratoire d'Astrophysique de Toulouse-Tarbes, Universit\'e de Toulouse, CNRS, 14 av. Edouard Belin, 31400 Toulouse, France\\
              \email{jballot@ast.obs-mip.fr}
         \and
             Department of Applied Mathematics, University of Sheffield, Hicks Building, Hounsfield Road, Sheffield S3 7RH, UK
             }

  \abstract
   {Using the Two-dimensional Oscillation Program (TOP), we have explored the effects of rapid rotation on gravity modes in polytropic stars. Coriolis force, centrifugal distortion as well as compressible effects have been taken into account. Thanks to our complete calculation, we have first studied the validity domain of perturbative methods and started to explore properties of these modes. We focus on $l=1$ in this analysis.}


   \maketitle

\section{Introduction}
Rapidly rotating stars are common objects on the main sequence. These stars can be strongly distorted by the centrifugal force, and some of them are known to spin almost as fast as their break-up frequency. Fast rotators are easily found in some classes of pulsating stars, such as $\delta$ Scuti, $\gamma$ Doradus, $\beta$ Cepheid or SPB (slow pulsating B) stars.
Acoustic (p) modes are excited in $\delta$ Sct and $\beta$ Cep stars, whereas gravity (g) modes are responsible for SPB and $\gamma$ Dor oscillations. Rotation effects cannot generally be neglected for frequency computation even for slow rotation. However, when the rotation frequency $\Omega$ is small relative to mode frequencies $\omega$ and to $\Omega_K=\sqrt{GM/R_{eq}^3}$, rotation effects can be treated as a perturbation ($M$ and $R_{eq}$ are the mass and equatorial radius of the star).
Such approaches ease frequency computations, but have a limited application range in term of rotation rate. \citet{Lignieres06,Reese06} have investigated the limits of perturbation theory for p-mode computation showing that taking the centrifugal distortion into account is needed to correctly predict mode frequencies.

In this paper we investigate the effects of rapid rotation on gravity mode frequencies by performing a full two-dimensional computation of modes in polytropic models of stars. The effects of rotation on the modes are fully taken into account.

\section{Methods}

\begin{figure}[htbp]
     \begin{center}
       \includegraphics[width=8cm]{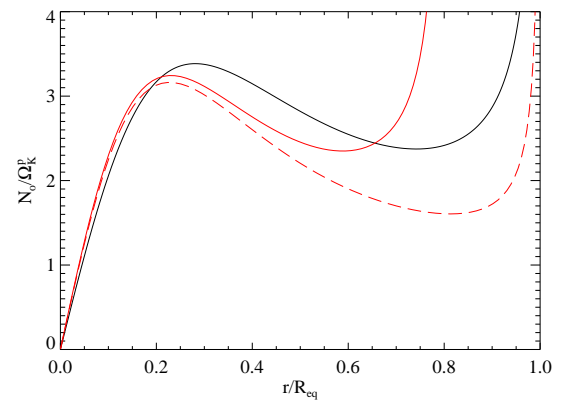}
     \end{center}
     \caption{Solid black line shows the profile of Brunt-V\"ais\"al\"a frequencies $N_o$ normalized by $\Omega_K^p$ for the non-rotating star. Red lines show $N_o$ for the model $\Omega=0.7\Omega_K$ along the polar (solid line) and equatorial radii (dashes).}
     \label{fig:BV}
   \end{figure}

We consider for this work polytropic models of stars with a polytropic index $N=3$ corresponding to a radiative zone.
The 2D structure of the rotating polytrope is computed as described in \citet{Lignieres06}.
We have considered models spinning uniformly at the rotation frequency $\Omega$ between $0$ and $0.7\Omega_K$.
Models are described in a spheroidal geometry \citep[more details for instance in][]{Reese06}. We have used a resolution of $n_r=96$ in the pseudo-radial direction and $L_{max}=32$ spherical harmonics in latitude. This resolution is sufficient enough to accurately model the centrifugal effects for the maximal value of $\Omega$ we have considered.
Figure~\ref{fig:BV} shows the profile of the Brunt-V\"ais\"al\"a frequency $N_o$ normalized by $\Omega_K^p=\sqrt{GM/R_{pol}^3}$ (where $R_{pol}$ is the polar radius) for the two extreme models ($\Omega$=0 and $0.7\Omega_K$).We see that $N_o$ does not change too much in terms of values and profiles for the different models.

We solve the eigenvalue problem corresponding to adiabatic small perturbations of the polytropic model of star. To do that, we used the Two-dimensional Oscillation Program (TOP) based on \citet{Reese06}.
The modes being decomposed on spherical harmonics $Y_l^m$, the resolution is given by the truncation of the decomposition. For this work, we have considered spherical harmonics up to $L=40+|m|$, i.e. 20 coupled harmonics. The radial discretization is the same as for the models, i.e. $n_r=96$.

\section{Results}
\begin{figure}[!ht]
     \begin{center}
       \includegraphics[width=8cm]{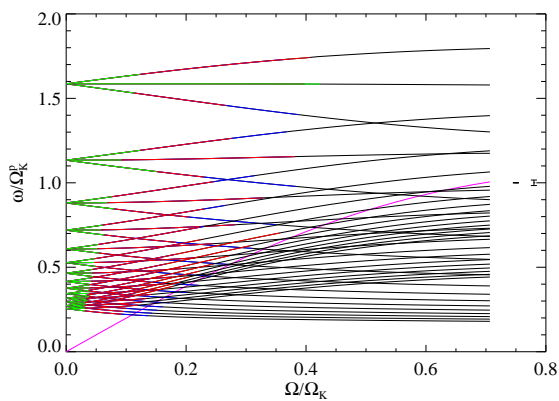}
     \end{center}
     \caption{Evolution of the frequencies of $l=1$ modes. Frequencies are computed in the co-rotating frame. Perturbative approximations have been tested for a star with $M=1.5M_{\sun}$ and $R_{pol}=1.6R_{\sun}$ and an error of $0.1\:\mu$Hz: green/red/blue parts of curves indicate that 1st/2nd/3rd order is sufficient to reproduce complete calculations. The magenta line indicates $\omega=2\Omega$.}
     \label{fig:valdom}
   \end{figure}

We have computed the solutions with $l=1$ ($m=-1,0,1$) and radial orders $n=-1...-14$ without rotation, and followed the evolution of the frequencies of these 42 modes by slowly increasing the rotation rate to $0.7\Omega_K$. These frequencies are plotted in Fig.~\ref{fig:valdom}.

From this data, we can compute the frequencies that would be determined in a perturbative approach. In such an approximation, frequencies are expressed as
\begin{equation}
\bar\omega_{n,l,m}^{pert}=\bar\omega_{n,l}+ C_{l,m}^1 \bar\Omega + C_{l,m}^2 \bar\Omega^2 + C_{l,m}^3 \bar\Omega^3 + {\cal O}(\bar\Omega^4),
\end{equation}
where $\bar\omega=\omega/\Omega_K^{p}$ and $\bar\Omega=\Omega/\Omega_K^{p}$. We have considered approximations to the 3rd order.
The term $\bar\omega_{n,l}$ is nothing else but the frequency without rotation, whereas the coefficients $C_{l,m}^i$ are numerical computations of the $i$th-derivative $d^i\bar\omega_{n,l,m}/d\bar\Omega^i$ at $\Omega=0$ obtained thanks to our full computation.
Since these perturbative coefficients contain all the considered physics, they are ideal coefficients. In a sense, any perturbative theory aims to approximate the coefficients we have found.
We have verified that the frequencies without rotation $\bar\omega_{n,l}$ are in perfect agreement ($10^{-8}$) with previous computations \citep{CDM94}. We have also compared $C_{l,m}^1$ to its analytical expression \citep{Ledoux51}, which gives the same results within relative errors of $10^{-9}$. Concerning the coefficients $C_{l,m}^2$ and $C_{l,m}^3$ we have estimated the errors to be around $10^{-5}$. This last error is deduced from the errors estimated for our numerical derivatives.

We have normalized frequencies by $\Omega_K^{p}$ since the polar radius is expected to be a slowly varying function of $\Omega$ in real stars, as opposed to $R_{eq}$.

We have then derived the validity domains for 1st, 2nd and 3nd order methods. These domains depend i) on the error we allow on the frequency and ii) on the frequency $\Omega_K^p$ that depends on the star.
Figure~\ref{fig:valdom} shows the validity domain of 1st, 2nd and 3nd order methods for a typical $\gamma$ Dor star ($M=1.55M_{\sun}$, $R_{pol}=1.6R_{\sun}$) for an error of 0.1$\:\mu$Hz. In this case, perturbative approaches can be sufficient to determined the higher frequencies as long as $\Omega \lesssim 0.35\Omega_K$ but only for modes with $\omega>2\Omega$. The domain of validity would be more extended for larger error tolerances or for more massive stars (i.e. higher $\Omega_K^p$) such as SPB stars. Nevertheless we observe in any case a strong departure from the exact frequencies as $\omega<2\Omega$. This can be interpreted as the fact that the volume of the resonant cavity is modified by the effects of the Coriolis force when $\omega<2\Omega$ \citep[see for instance][]{Dintrans00}.

   \begin{figure}[!ht]
     \begin{center}
       \includegraphics[width=8cm]{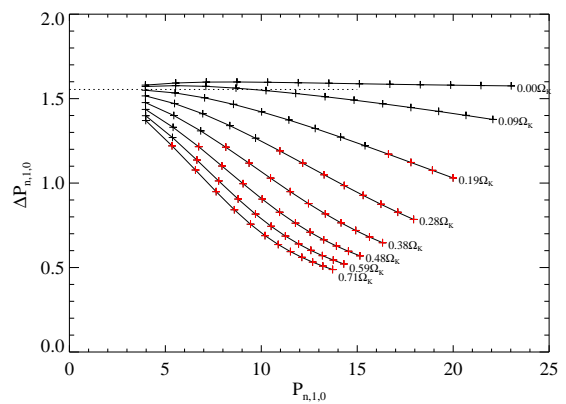}
     \end{center}
     \caption{Period spacing $\Delta P_{n,l}$ for $l=1$ and $m=0$ as a function of the mode periods for different rotation speeds. Red crosses indicates sub-inertial modes. Dots indicate the asymptotic value from \citet{Tassoul80}. Periods are in unit of $2\pi/\Omega_K^p$.}
     \label{fig:per}
   \end{figure}

For a non-rotating star, the period spacing $\Delta P_{n,l}=P_{n+1,l,0}-P_{n,l,0}$, where the period $P_{n,l,m}=2\pi/\omega_{n,l,m}$, is known to be --asymptotically-- independent of $n$, in the absence of strong structure gradients \citep{Tassoul80}. Figure~\ref{fig:per} shows the evolution of this spacing, for $l=1, m=0$ modes when rotation increases. While this spacing is almost constant at low rotation, even for low-order g modes, as soon as the rotation becomes non negligible, it is no more the case and the period spacing follows a more complex law. As a consequence, at higher rotation rates, Tassoul's formula cannot be used for mode identification.

\section{Conclusions}
This short report shows some preliminary results obtained on g modes. This work has been extended to $l=2$ and $3$ modes and will be presented in more details in an forthcoming paper (Ballot et al., in preparation).

\begin{acknowledgements}
The authors acknowledges support through the ANR project Siroco. Many of the numerical calculations were carried out on the supercomputing facilities of CALMIP (``CALcul en MIdi-Pyr\'en\'ees'') which is gratefully acknowledged.	
\end{acknowledgements}

\bibliographystyle{aa}
\bibliography{biblio}

\begin{thebibliography}{6}
\expandafter\ifx\csname natexlab\endcsname\relax\def\natexlab#1{#1}\fi

\bibitem[{{Christensen-Dalsgaard} \& {Mullan}(1994)}]{CDM94}
{Christensen-Dalsgaard}, J. \& {Mullan}, D.~J. 1994, \mnras, 270, 921

\bibitem[{{Dintrans} \& {Rieutord}(2000)}]{Dintrans00}
{Dintrans}, B. \& {Rieutord}, M. 2000, \aap, 354, 86

\bibitem[{{Ledoux}(1951)}]{Ledoux51}
{Ledoux}, P. 1951, \apj, 114, 373

\bibitem[{{Ligni{\`e}res} {et~al.}(2006){Ligni{\`e}res}, {Rieutord}, \&
  {Reese}}]{Lignieres06}
{Ligni{\`e}res}, F., {Rieutord}, M., \& {Reese}, D. 2006, \aap, 455, 607

\bibitem[{{Reese} {et~al.}(2006){Reese}, {Ligni{\`e}res}, \&
  {Rieutord}}]{Reese06}
{Reese}, D., {Ligni{\`e}res}, F., \& {Rieutord}, M. 2006, \aap, 455, 621

\bibitem[{{Tassoul}(1980)}]{Tassoul80}
{Tassoul}, M. 1980, \apjs, 43, 469

\end{thebibliography}
\end{document}